\documentclass[12pt]{article}

\usepackage{epsfig}
\usepackage{psfrag}
\usepackage{latexsym}
\usepackage[DIV13]{typearea}
\usepackage{amsmath}
\usepackage{amssymb}
\usepackage{amsfonts}
\usepackage{bbold}
\usepackage[footnotesize]{caption2}
\usepackage{graphicx}
\usepackage[center,footnotesize,hang]{subfigure}
\usepackage{url}
\usepackage{color}
\usepackage{cite}
\usepackage{pstricks}
\usepackage{inputenc}
\usepackage{pst-coil}
\usepackage{verbatim}
\usepackage{bbm}
\newpsobject{grilla}{psgrid}{subgriddiv=1,griddots=10,gridlabels=6pt}

\def\5{\overline 5}

\newcommand{\diag}{\text{diag}}

\newcommand{\unity}{\mathbb{1}}
\newcommand{\nn}{\nonumber}

\def\eq#1{{eq.~(\ref{#1})}}

\def\vev#1{\left\langle #1\right\rangle}

\def\hbar{\hspace{0pt}\raisebox{1pt}{$-$} \hspace{-7pt} h}

\newcommand{\beq}{\begin{equation}}
\newcommand{\eeq}{\end{equation}}
\newcommand{\bac}{\beq\begin{array}}
\newcommand{\eac}{\end{array}\eeq}
\newcommand{\ba}{\begin{array}}
\newcommand{\ea}{\end{array}}
\newcommand{\bea}{\begin{eqnarray}}
\newcommand{\eea}{\end{eqnarray}}

\newcommand{\bmat}{\begin{pmatrix}}
\newcommand{\emat}{\end{pmatrix}}

\begin{document}

\begin{center}
{\Large\bf Tri-Permuting Mixing Matrix and  predictions for $\theta_{13}$ }

\end{center}

\begin{center}
 \vskip 0.3 cm
{\large Federica Bazzocchi}\footnote{e-mail address: fbazzo@sissa.it},\\
\vskip .2cm
SISSA and INFN, Sezione di Trieste, \\
via Bonomea 265, 34136 Trieste, Italy
\vskip .1cm

\end{center}
\vskip 0.7cm
\begin{abstract}
We introduce a new texture for neutrino mixing  named Tri-Permuting  (TP) mixing matrix. This pattern is characterized by maximal solar and atmospheric angles and by a large reactor angle satisfying  $\sin \theta_{13}=1/3$. The correct lepton mixing matrix is obtained when combining the charged lepton mixing matrix with the neutrino one. In this way  we get new predictions for $\theta_{13}$ with respect to those obtained by the well studied TBM or BM mixing patterns. We present a specific model that gives rise to TP  mixing in the neutrino sector as well as the  required corrections  from  the charged lepton one.
\end{abstract}
\setcounter{footnote}{0}
\vskip2truecm


\section{Introduction}

 The most recent T2K\cite{Abe:2011ks} and MINOS \cite{MINOS}  results  have shown at 2.5 and 1.7 $\sigma$  respectively  the evidence of a $\theta_{13}\neq 0$ in the lepton mixing matrix.  The first global analysis\cite{Fogli:2011qn}   have confirmed their results  giving at $3\sigma$ level the range $0.001\leq\sin \theta_{13}^2 \leq 0.044$ ($0.005\leq\sin \theta_{13}^2 \leq 0.050$) for the NH (IH) case. Comparable results have been obtained by the most recent global analysis\cite{Schwetz:2011zk}  that  have slightly lowered the upper bound  $0.001\leq\sin \theta_{13}^2 \leq 0.035$ ($0.015\leq\sin \theta_{13}^2 \leq 0.039$) for the NH (IH) case.
 
 While waiting for more statistics and forthcoming tests of these results neutrino phenomenology community have showed an impressive fast and conspicuous productivity in proposing new textures and models that could account for the correct $\theta_{13}$ size. The majority of these analysis have been devoted to a re-consideration of the possible  corrections to TriBiMaximal (TBM) mixing. TBM pattern predicts at  leading order (LO) a vanishing  $\theta_{13}$. Even  in the first TBM predicting models\cite{Altarelli:2010gt}  a non vanishing $\theta_{13}$ was indeed predicted at next to leading order (NLO) typically quite small, of order   $\theta_C^2$, with $\theta_C\sim.23$ the Cabibbo angle. In the last months different scenarios based mainly on discrete symmetries have been proposed to modify TBM texture and predicting a $\theta_{13}\neq 0$ up to 10  degrees\cite{allTBM, Marzocca:2011dh}. Other possibilities have been considered in \cite{other}.

Even before T2K and MINOS recent data a  promising idea  to get a $\theta_{13}\neq 0$ was given by BiMaximal (BM) mixing.
  
  In the context of BM mixing the basic idea is that at LO solar and atmospheric angles are maximal and the reactor angle is zero\cite{BMmixing,AFM_BimaxS4,Meloni:2011fx}.
Then at next leading order (NLO)  only the solar and the reactor angles  get corrections of order the Cabibbo angle  $ \theta_C$ while the atmospheric keeps unchanged. Finally at next-next leading order (NNLO) even the atmospheric angle may get corrections but these are of order $\theta_C^2$ thus allowing to fall in the experimental data range. In this picture NLO corrections at the lepton mixing matrix  arise by diagonalizing  the charged lepton sector.  The reason is  seeded in the original motivation to study BM mixing, that  is  quark-lepton complementarity\cite{Complementarity}.  We remind that recently it has been  shown how a relative large $\theta_{13}$ may  arise from the charged lepton sector  in the context of  SU(5)  assuming exact  TBM mixing in the neutrino sector\cite{Marzocca:2011dh}.

At the moment while we are looking for more statistics and new analysis  to confirm and deline the  $\theta_{13}$ range  one of the   challenges is finding a texture that has a non vanishing $\theta_{13}$ at LO--eventually even  too large--and thus making it smaller thanks to adequate  corrections\cite{Toorop:2011jn}.

At the light of the most recent results there is an intrinsic tension in the BM  assumption. Consider the BM choice
\beq
\theta_{12}=\theta_{23}= -\frac{\pi}{4}\,, \quad \theta_{13}=0\,,
\eeq
and assume that 
the charged lepton mass matrix is diagonalized on the left by a rotation in the 12 sector of order the Cabibbo angle parametrized as
\beq
\label{Uch}
U_{e}= \left( \begin{array}{ccc} \cos \theta &\sin \theta e^{i \delta}&0\\ -\sin \theta e^{i \delta}& \cos \theta&0\\0&0&1 \end{array}\right)\,,
\eeq
where $\delta$ is a possible CP Dirac phase.

\begin{figure}[h]
\begin{center}
\includegraphics[width=4in]{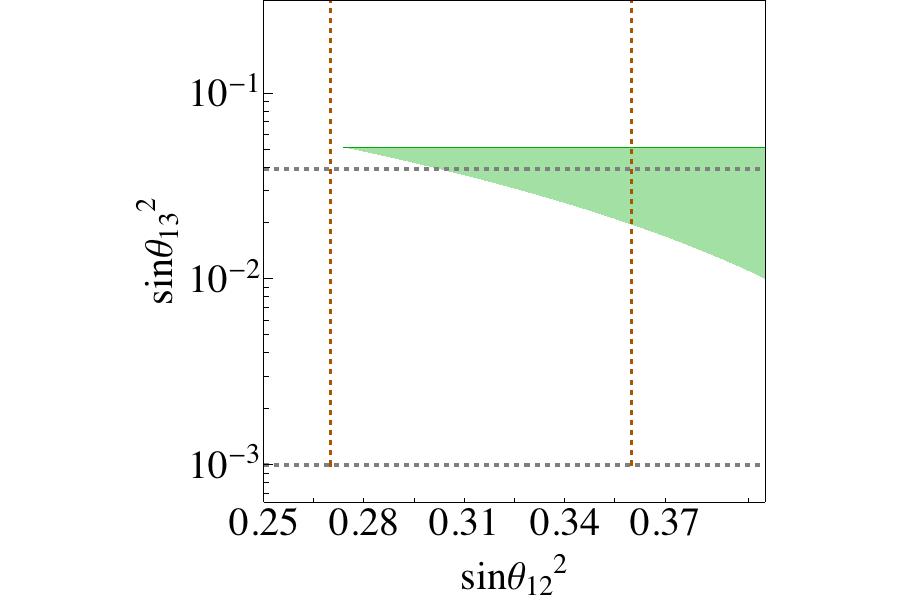} 
\caption{\it The reactor  versus the solar angle in the case of the BM mixing matrix corrected by a rotation in the 12 plane of the charged lepton mass matrix. Vertical and horizontal lines bound the $3\sigma$ range for $\sin\theta_{12}^2$ and $\sin\theta_{13}^2$ respectively.} \label{parBM}
\end{center}
\end{figure}

Then
one finds that the lepton mixing angles are given by
\bea
\label{BMex}
\theta_{23}  &\sim&- \frac{\pi}{4} +  \frac{1}{4} \theta^2+\mathcal{O}(\theta^3)\,,\nn\\
 \theta_{12}&\sim&-\frac{\pi}{4}- \frac{1}{\sqrt {2}}\theta \cos \delta +\mathcal{O}(\theta^3)\,,\nn\\
  \theta_{13}&\sim & \frac{1}{\sqrt{2}}\theta\,.
\eea
The solar angle wants a \emph{large}  $\theta$ of order the Cabibbo angle $\theta_C$, while the most recent fit indicates that $\theta_{13}$ is large but not too much. The results is shown in fig. \ref{parBM}: according to the simple parametric expansion of \eq{BMex} the BM $\theta_{13}$ prediction is large and could be ruled out by an improvement of precision that could low the $3\sigma$ upper bound. However in more realistic scenarios the values allowed for $\theta_{13}$ are spreaded, but we still may conclude that if the upper $3\sigma$ limit on $\theta_{13}$ would be lowered the BM pattern would be strongly disfavored.

\begin{figure}[h]
\begin{center}
\includegraphics[width=4in]{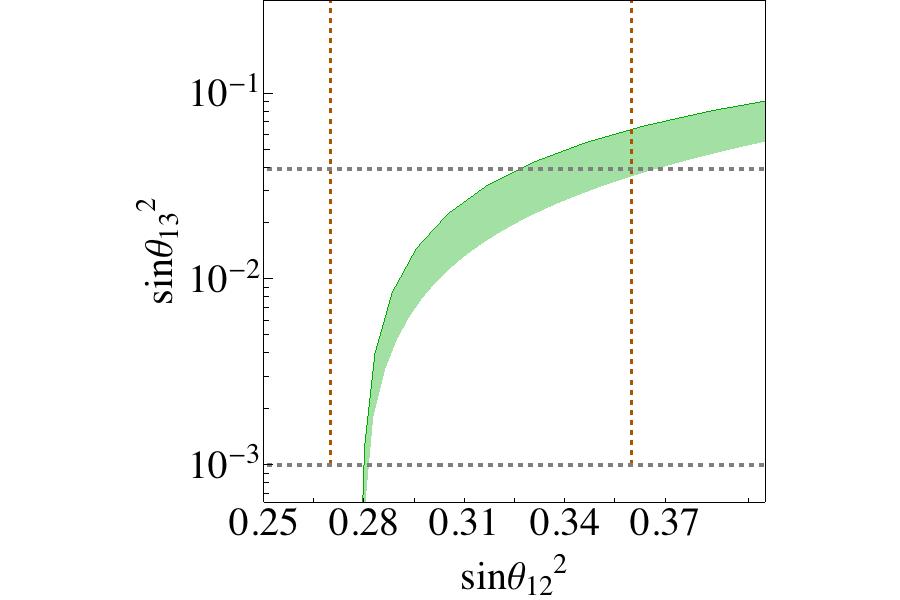} 
\caption{ \it The reactor  versus the solar angle in the case of the TP mixing matrix corrected by a $12$ rotation in the charged lepton sector . Vertical and horizontal lines bound the $3\sigma$ range for $\sin\theta_{12}^2$ and $\sin\theta_{13}^2$ respectively.} \label{parTP}
\end{center}
\end{figure}

In this paper we present a complementary picture to that offered by the BM pattern that we define as  the Tri-Permuting  (TP) mixing  matrix. The name reminds that the 3 eigenvectors are identical up to permutations and change of signs. This TP mixing  matrix is defined by two maximal angles and a large $\theta_{13}$ according to

\beq
\sin \theta_{12}=\sin \theta_{23}= -\frac{1}{\sqrt{2}}\,, \quad \sin \theta_{13}=\frac{1}{3}\,.
\eeq

\begin{equation}\label{ULO}
U_{TP}\sim
\frac{1}{3}\left(
\begin{array}{ccc}
2&-2&1\\
2&1&-2\\
1&2&2
\end{array}
\right)\,.
\end{equation}
Under the previous assumption that  charged lepton mass matrix  is diagonalized through the rotation given in \eq{Uch} we get
\bea
\label{TPang}
\theta_{23}  &\sim&-\frac{ \pi}{4} +  \frac{1}{4}\theta \cos\delta\,+ \frac{1}{16}\theta^2( 4+\cos 2 \delta)+\mathcal{O}(\theta^3)\,,\nn\\
 \theta_{12}&\sim&-\frac{\pi}{4}- \frac{3}{4}   \theta \cos\delta-\frac{ 3}{16}\theta^2 \cos 2 \delta+\mathcal{O}(\theta^3)\,,\nn\\
  \theta_{13}&\sim&\frac{1}{3} +\frac{2}{3} \theta \cos \delta  -\frac{1}{6}\theta^2( 1+4 \sin  \delta^2)+\mathcal{O}(\theta^3)\,.
\eea
Notice that while $\theta_{12}$ and $\theta_{13}$ receive a correction of order $\sim \theta \cos\delta$, $\theta_{23}$ is corrected by $\theta \cos \delta /4\sim \theta_C^2$ if $\theta \sim \theta_C$ the Cabibbo angle. Moreover  to constrain $\theta_{12}$ in the correct range we need $\cos \delta<0$ that gives a correction to $\theta_{13}$ in the right direction. This is explicitly shown in fig. \ref{parTP}. At the same time we get a prediction for the Dirac CP phase that in this scenario is given by
\beq
\label{expdelta}
\delta_l\sim 2( \theta \sin\delta -\theta^2 \sin 2 \delta)\,.
\eeq

In the next section we  introduce the framework in which TP mixing matrix arises. In sec.\ref{model} we  build a renormalizable model that provides such a texture. Neutrino phenomenological implication are discussed  in sec.\ref{anal}  and sec.\ref{conc} is devoted to our conclusions.

\section{Residual symmetries}
\label{res}
 
It is well known that  under the assumption that neutrinos are majorana particles if there is any residual symmetry behind neutrino mass matrix this is at most a $Z_2 \times Z_2$  flavor symmetry \cite{Feruglio:2011qq,Toorop:2011jn}. While it is clear how the $Z_2 \times Z_2$ acts on the neutrino mass eigenstates since the  three mass eigenstates must  have \emph{flavor parity}  (+,+),(+,-),(-,-), it is an open question how it acts on the  neutrino interaction eigenstates.
In the most general case given the three left handed neutrinos  $\nu_L\sim(\nu_{L_1},\nu_{L_2},\nu_{L_3})$  each $Z_2$ flavor  symmetry $S_{1,2}$, $S_{1,2}^2=I$, acts on  $\nu_L$ as 
\beq
\nu_L \to S_i \nu_L
\eeq
and it holds that $[S_1,S_2]=0$. Thus the neutrino mass matrix may be written in terms of the three eigenvectors of $S_1$ and $S_2$, $v_i$,  satisfying 
\bea
S_1 v_1= v_1  &\quad &S_2 v_1 =v_1\,,\nn\\
S_1 v_2=v_2 &\quad& S_2 v_2=-v_2 \,, \nn\\
S_1 v_3=-v_3 &\quad & S_3 v_3 =-v_3\,.
\eea
As consequence the effective light neutrino  majorana mass matrix may be written as
\beq
m_\nu= m_1 v_1^T v_1+ m_2 v_2^T v_2 +m_3 v_3^T v_3\,.
\eeq
This approach has been used in many scenarios and typically addressed as sequential dominance\cite{SD}.  Determining $S_1$ and $S_2$ fixes uniquely the lepton  mixing matrix if the charged lepton mass matrix is  diagonal.  $S_1$ and $S_2$ corresponding to the TP mixing matrix are given by

\bea
S_1=\left(
\begin{array}{ccc}
 \frac{7}{9} & \frac{4}{9} & -\frac{4}{9} \\
 \frac{4}{9} & \frac{1}{9} & \frac{8}{9} \\
 -\frac{4}{9} & \frac{8}{9} & \frac{1}{9}
\end{array}
\right)\,,&\quad & S_2 =\left(
\begin{array}{ccc}
 -\frac{1}{9} & \frac{8}{9} & \frac{4}{9} \\
 \frac{8}{9} & -\frac{1}{9} & \frac{4}{9} \\
 \frac{4}{9} & \frac{4}{9} & -\frac{7}{9}
\end{array}
\right)\,.
\eea

A diagonal charged lepton mass matrix  is invariant under  an infinite choice of abelian symmetries  since charge assigments for the left handed fields may be compensated by the  corresponding  right ones.   A natural   choice is given by $Z_e\times Z_\mu\times Z_\tau$.  Clearly this symmetry   has to be broken by  soft terms    or by  NLO  contributions if the correction to the TP mixing matrix has to arise by the charged lepton sector.

\section{The model}
\label{model}

In this section we build a renormalizable model that provides the TP mixing matrix.  We assume that no other heavy matter fields exist a part from those we report in tab. \ref{tab:fields}, thus the Yukawa lagrangian we write in \eq{Yuk} is complete and no NLO terms have to be taken into account. The model is based on the  flavor symmetry $G_f\sim SU(3)_F\times U(1)_F$ and matter and  scalar fields charge assigments are reported in tab.\ref{tab:fields}.  Left handed doublets transform as a triplet of $SU(3)_F$. Standard model (SM) right handed charged lepton are $SU(3)_F$ singlet and  charged under $U(1)_F$. Among the matter fields we have 2 right handed neutrinos, singlet under $SU(3)_F\times U(1)_F$,  a vectorial couple of heavy SM singlets $\Sigma$,$\overline{\Sigma}$, transforming as 3 and $\overline{3}$ respectively under $SU(3)_F$, another  vectorial couple of heavy SM $SU(2)$  singlet charged under $U(1)_Y $, $F$, $F^c$,  3 and $\overline{3}$  of $SU(3)_F$ respectively. We introduce five $\overline{3}$ scalar fields, three of them are charged under $U(1)_F$. In addition   we impose an extra  $Z_2$ symmetry under which all the fields are  even with the exception of  one right handed neutrino, $\nu^c_2$,  and  one scalar triplet, $\phi_2$, that are odd.

\begin{table}[!h]
\begin{center}
\begin{tabular}{|c||c|c|c|c|c|c||c|c|c|c||c|c|c|c|c|c|}
\hline
  &&&&&&&&&&&&&&&& \\
  Matter & $L$ & $\nu^c_1$&$\nu^c_2$& $e^c$ & $\mu^c$& $\tau^c$& $\Sigma$ &$\bar{\Sigma}$&$ F$ &$F^c$ &$\phi_1$&$\phi_2$&$\phi_e$&$\phi_\mu$&$\phi_\tau$&$H$ \\
  &&&&&&&&&&&&&&&& \\[-0,3cm]
  \hline
   &&&&&&&&&&&&&&&& \\
  $SU(2)_L$ & 2& 1&1&1&1&1&1&1&1&1&1&1&1&1&1&2\\
   &&&&&&&&&&&&&&&&  \\[-0,3cm]
    $U(1)_Y$ & -1/2& 0&0&1&1&1&0&0&-1&1&0&0&0&0&0&1/2\\
   &&&&&&&&&&&&&&&&  \\[-0,3cm]
  $SU(3)_F$ & $3$ & $1$ & $1$& $1$ &1&1&3&$\overline{3}$&3&$\overline{3}$ &$\overline{3}$ &$\overline{3}$ &$\overline{3}$ &$\overline{3}$ &$\overline{3}$&1 \\
  &&&&&&&&&&&&&&&& \\[-0,3cm]
  $U(1)_{F}$ & 0 & 0 & 0 &$\alpha$&$\beta$ &$\gamma$&0&0&0&0&0 & 0 &$-\alpha$&$-\beta$ &$-\gamma$& 0 \\
   &&&&&&&&&&&&&&&& \\
    \hline
  \end{tabular}
\end{center}
\caption{\it Transformation properties of the matter fields. The choice of the charged lepton  $U(1)_F$ charges is arbitrary once the condition $\alpha\neq \beta\neq \gamma$ is satisfied.}
\label{tab:fields}
\end{table}

\subsection{Mass matrices}

Given the field content of tab.\ref{tab:fields} the lagrangian reads as
\bea
\label{Yuk}
\mathcal{L}&=& k L H \overline{\Sigma}+ (y_1 \Sigma \phi_1+y'_1 \overline{\Sigma} \phi^*_1) {\nu}^c_1+(y_2 \Sigma \phi_2+y'_2 \overline{\Sigma} \phi^*_2) {\nu^c}_2+ M_\Sigma \Sigma  \overline{\Sigma}+ \frac{M_1}{2} \nu^c_1 \nu^c_1+\frac{M_2}{2} \nu^c_2 \nu^c_2\nn\\
&+&y_F L \tilde{H} F^c+ y_e F \phi_e e^c+ y_\mu F \phi_\mu \mu^c+ y_\tau F \phi_\tau \tau^c+ M_F F F^c\,,
\eea
where $\tilde{H}=i \sigma_2 H$ with $H$ the usual SM higgs doublet and  $\sigma_2$ the Pauli matrix. We have omitted $SU(2)$ indices to simplify the notation. As already stated  this is the full Yukawa lagrangian  thus no NLO corrections have to been included. 

In sec.\ref{res} we have said that TP mixing matrix is obtained when neutrino and charged lepton  mass matrices are invariant under $Z_2 \times Z_2$ and $Z_e\times Z_\mu \times Z_\tau$ respectively. This means that the flavor group $G_f$ has to be broken following different patterns in the neutrino and charged lepton sector as it usually happens in the context of discrete flavour symmetries\cite{Altarelli:2010gt}.  In our scenario this is realized when the five scalars $\phi_i$  develop vacuum expectation values (VEVs) as
\bea
\label{vev}
\vev{\phi_1}\sim(2,2,1)\,, & \quad& \vev{\phi_2}\sim(2,-1,-2)\,, \nn\\
\vev{\phi_e}\sim(1,0,0)\,, & \quad& \vev{\phi_\mu}\sim(0,1,0)\,, \quad \vev{\phi_\mu}\sim(0,0,1) \,.\nn\\
\eea
In   sec.\ref{pot}  we sketch how this alignment may be realized.

When flavor and electroweak symmetries are broken the neutrino mass matrix is a 5 block matrix  that in the basis $(\nu_L, \overline{\Sigma},\Sigma,\nu^c_1,\nu^c_2)$ is given by
\bea
\label{M1}
M_\nu&=&   \left( \begin{array}{ccc} 0& m_D & 0\\
m_D^{T} & M_0  &\lambda \\
0&\lambda& M_{\nu^c}
\end{array}\right)\,.
\eea
with
\bea
m_D=( k v_h \cdot \unity ,0)\,, &\quad&
\lambda=\left( \begin{array} {cc} \lambda'_1&\lambda'_2\\
\lambda_1&\lambda_2\\ \end{array}\right)\,,\nn\\
M_{\nu^c}=Diag(M_1,M_2)\,, &\quad&M_0=\left( \begin{array} {cc} 0&M_\Sigma\cdot \unity \\
M_\Sigma\cdot \unity &0\\ \end{array}\right)\,,\nn\\
\eea
being $\unity$ the identity $3\times 3$ matrix.
The $\lambda$'s are defined as
\bea
\lambda_1,\lambda'_1= y_1,y_1' v_1\left( \begin{array}{c} 2\\2 \\1\end{array} \right)&\quad & \lambda_2,\lambda'_2= y_2,y_2' v_2\left( \begin{array}{c} 2\\-1 \\-2\end{array} \right)
\eea
Under the assumption $M_\Sigma> M_{1,2}>\lambda_{1,2},\lambda_{1,2}'$
it is convenient defining the  spinor  $\Sigma_1$ and ${\Sigma}_2$ 
\beq
\Sigma_1 =\frac{1}{\sqrt{2}}(\overline{\Sigma}+\Sigma) \quad \Sigma_2 =\frac{1}{\sqrt{2}}(-\overline{\Sigma}+\Sigma) \,.
\eeq
In this way \eq{M1}   becomes
\bea
\label{M2}
M_\nu&=& \left( \begin{array}{ccccc} 0& \tilde{m}_D & 0\\
\tilde{m}_D^T &\tilde{M}_0& \tilde{\lambda}\\
0& \tilde{\lambda}^T& M_{\nu^c}
\end{array}\right)\,.
\eea
$M_\nu$ may be  sequentially diagonalized by using  the block diagonalization  method introduced in \cite{Schechter:1980gr}.
First the method is applied to the block involving the heavy fields $( {\Sigma}_1,\Sigma_2,\nu^c_1,\nu^c_2)$. The unitarity matrix that diagonalize the block is defined as
\beq
U_H=\sim \left(\begin{array}{cc}1-\frac{B B^T}{2}& B\\-B^T &1-\frac{B^T B}{2} \end{array} \right)
\eeq
with 

\beq
B\sim -\tilde{M}_0^{-1} \tilde{\lambda}\,.
\eeq
The lightest singlet neutrinos mass matrix becomes
\beq
\tilde{M}_{\nu^c}=M_{\nu^c}-\tilde{\lambda}^T \tilde{M}_0^{-1}\tilde{\lambda}\,.
\eeq
The effective light  neutrino mass matrix is given by  the usual type I see saw formula according to \bea
m_\nu& \sim& -\tilde{m}_D B\frac{1}{ \tilde{M}_{\nu^c}} B^T m_D^T \sim-\tilde{m}_D\frac{1}{ \tilde{M}_0} \tilde{\lambda}\frac{1}{ {M}_{\nu^c}} \tilde{\lambda} ^T \frac{1}{\tilde {M}_0^{T}} m_D^T\,,
\eea
and presents the form
\beq
\label{massnu}
m_\nu=\left(\begin{array}{ccc}  4 x + 4 y& 4 x -2 y& 2 x -4 y\\ 4 x -2 y& 4 x +y& 2 x+ 2 y\\ 2 x -4 y& 2 x+2 y& x +4 y \end{array} \right)
\eeq
with 
\beq
x= 2 k^2 v_h^2 \frac{ y_1^2 v_1^2}{M_1 M_\Sigma^2}\,,\quad y= 2 k^2 v_h^2  \frac{ y_2^2 v_2^2}{M_2 M_\Sigma^2}\,.
\eeq
 The previous $m_\nu$ is diagonalized by  $U_{TP}$ with eigenvalues $( 9 x, 9y,0)$. Thus our realization allow only the IH spectrum.

For what concerns the charged lepton sector in addition to the  SM fields we have the heavy fields $F,F^c$.  When the vacuum alignment coincides exactly with that in  \eq{vev} the full charged lepton left-right mass matrix    presents a  trivial block structure
\beq
\label{chl}
M_{ch}=\left( \begin{array}{cc} 0&y_F v_H \cdot \unity \\Y& M_F \cdot \unity  \end{array}\right)\,,
\eeq
with $Y=\mbox{Diag}(y_e v_e,y_\mu v_\mu,y_\tau v_\tau)$. By integrating out the heavy fields the SM charged lepton mass matrix is diagonal and
\beq
(m_e,m_\mu ,m_\tau)=\frac{y_F v_H}{M_F} (y_e v_e,y_\mu v_\mu,y_\tau  v_\tau)\,.
\eeq
In sec.\ref{pot}  it is discussed how the alignments given in \eq{vev} get corrections due to  the presence of soft terms needed to give mass to the Goldstone Bosons (GBs)  arising by minimizing the potential. Specifically, for what concerns $\phi_{e,\mu,\tau}$, \eq{vev} have to be substituted by
\beq
\vev{\phi_{e,\mu,\tau}} + \epsilon_{e,\mu,\tau} (2,2,1)\,.
\eeq
In this way the  block $ Y $ is substituted by 
\beq
\tilde{Y}=\left(  \begin{array}{ccc} y_e( v_e + 2 \epsilon_e) & 2  y_\mu  \epsilon_\mu&2  y_\tau \epsilon_\tau \\ 2 y_e \epsilon_e & y_\mu( v_\mu + 2 \epsilon_\mu) &  2  y_\tau \epsilon_\tau\\ y_e \epsilon_e & y_\mu  \epsilon_\mu& y_\tau( v_\tau + 2 \epsilon_\tau)   \end{array}\right)\simeq  \left(  \begin{array}{ccc} y_e  v_e & 2  y_\mu  \epsilon_\mu&2  y_\tau \epsilon_\tau \\ 2 y_e \epsilon_e & y_\mu v_\mu &  2  y_\tau \epsilon_\tau\\ y_e \epsilon_e & y_\mu  \epsilon_\mu& y_\tau v_\tau  \end{array}\right)\,,
\eeq
since $\epsilon_x <<  v_x$.  Neglecting  terms proportional to $y_e v_e <<y_\mu v_\mu,y_\tau  v_\tau$ the charged lepton mass matrix squared presents the following structure
\beq
\label{mchl}
m_{ch}m^\dag_{ch}\simeq |m_\tau|^2 \left( \begin{array}{ccc}0& \epsilon'_\mu \frac{|y'_\mu|^2}{|y'_\tau|^2}& \epsilon'_\tau\\  {\epsilon'}^{*}_\mu \frac{|y'_\mu|^2}{|y'_\tau|^2}&\frac{|y'_\mu|^2}{|y'_\tau|^2}&  \epsilon'_\tau + {\epsilon'}^{*}_\mu \frac{|y'_\mu|^2}{|'y_\tau|^2} \\{ \epsilon'}_\tau^* &  {\epsilon'}^{*}_\tau + \epsilon'_\mu \frac{|y'_\mu|^2}{|y'_\tau|^2}&1 \end{array}\right)\,,
\eeq
 where we have defined  $y_x' = y_x v_x, \epsilon_x'= y_x v_x \epsilon_x/|m_\tau|^2$.  For construction  $\epsilon_x'<<1$ and  $| \epsilon'_\tau| <|\epsilon'_\mu|$ to fit the correct  ratio $|m_\mu|/|m_\tau|$ thus \eq{mchl} is diagonalized by
 \beq
U_e=\left( \begin{array}{ccc}1& \epsilon'_\mu& \epsilon_\tau'\\ {\epsilon'}^*_\mu&1&  \epsilon_\tau'\\
 {\epsilon'}^*_\tau&  {\epsilon'}^*_\tau &1 \end{array}\right)\,.
\eeq
The final lepton mixing matrix has  the desired structure and  in first approximation it is given by
\beq
\label{Ulep}
U_{lep}=U_e^\dag U_{TP} \simeq \left(
\begin{array}{ccc}
 \frac{2}{3}-\frac{2 \epsilon'_\mu }{3} & -\frac{\epsilon'_\mu }{3}-\frac{2}{3} &
   \frac{2 \epsilon'_\mu }{3}+\frac{1}{3} \\
 \frac{2 {\epsilon'}^*_\mu }{3}+\frac{2}{3} & \frac{1}{3}-\frac{2
   {\epsilon'}^*_\mu }{3} & \frac{{\epsilon'}^*_\mu }{3}-\frac{2}{3}
   \\
 \frac{1}{3} & \frac{2}{3} & \frac{2}{3}
\end{array}
\right)\,,
\eeq
where we have used  $| \epsilon'_\tau| <|\epsilon'_\mu|$.

\subsection{Vacuum alignment}
\label{pot}

Model based on flavor symmetries spontaneously broken in different directions in the charged lepton and neutrino sectors    have always to face off the problem of realizing the correct vacuum alignments. This affects TBM as well as BM models based on both discrete   and  continuous symmetries. The formers tend to break   in  one direction and therefore different techniques have been developed to break them in two directions. For the latter the situation is even worse since continuous symmetry do not develop a preferred direction to be broken  to and the minimum conditions   present an infinite degeneracy. For this reason in model based on  flavor continuous groups such as $SU(3)$ or $SO(3)$  the correct vacuum alignment is typically obtained by introducing soft breaking terms of the continuous symmetry.  These softs preserve an appropriate   discrete subgroup of the continuos symmetry  and through the minimization of the potential they select the correct directions \cite{deMedeirosVarzielas:2005ax}. Here we use the same approach.

In our model we have five $\overline{3}$ of the flavour group $G_f\sim SU(3)_F\times U(1)_F \times Z_2$. For five $\overline{3}$ of a generic $SU(3)$  the most generic potential is written as
\beq
V[\phi_i]= \mu^2_{ij} \phi_i^\dag \phi_j + \lambda_{ijkl} (\phi_i^\dag \phi_j) (\phi_k^\dag \phi_l)\,,
\eeq
where the sum other all the fields is understood.
In our case,  the abelian symmetries forbid many terms and in particular the scalar $SU(3)_F$ invariant potential has an accidental enlarged symmetry $SU(3)\times U(3)^3$: the first  $SU(3)$ involves $\phi_1$ and $\phi_2$  and one $U(3)$  per $\phi_i$, $i=e,\mu,\tau$. The generic vacuum configuration constrains the absolute  scalar VEVs
\beq
\vev{\phi_1}\,, \quad \vev{\phi_2} \,, \quad \vev{\phi_{e,\mu,\tau}}\,,
\eeq
and the accidental $SU(3)\times U(3)^3$ is completely broken giving rise to $8+9 \times 3=35$ GBs. 
This is a situation quite common in flavor models: the inclusion of the soft terms is needed both to trigger the correct alignments  and to give  mass to the  unwanted GBs. 

In our context we need three set of soft terms:
\begin{itemize}
\item[-] $V^1_{soft}$: it   triggers the correct vacuum alignment  for  $\phi_{e,\mu,\tau}$ and also breaks the accidental $U(3)^3$ symmetry  to $SU(3)$, giving mass to 19 GBs. At this level we are left with an accidental global symmetry $SU(3)\times SU(3)$
\item[-] $V^2_{soft}$: it triggers the correct vacuum alignment  for  $\phi_{1,2}$ breaking  $SU(3)\times SU(3)$ to $SU(3)$ and giving mass to other 8 GBs.
\item[-] $V^3_{soft}$: it  breaks the residual $SU(3)$ and gives mass to the last 8 GBs.
 \end{itemize}

A suitable example for $V^1_{soft}$ is given by
\beq
\label{V1s}
V^1_{soft}=[ m^2_{e\mu} (\phi_e^\dag \phi_\mu)+m^2_{\mu\tau} (\phi_\mu^\dag \phi_\tau)+  A \phi_e \phi_\mu\phi_\tau] +H.c.
\eeq
The basic assumption is that the soft terms slightly modify the first derivative system that in first approximation may be considered unchanged. Thus the first derivative system of the potential involving $\phi_{e,\mu,\tau}$ fixes $\vev{\phi_{e,\mu,\tau}}$ and we have always the freedom to choose one direction, for example
\beq
\vev{\phi_{e}}=v_e (1,0,0)\,.
\eeq
If  $m^2_{e\mu},m^2_{\mu\tau} >0$ and $A<0$ the  quadratic and quartic terms of \eq{V1s} select orthogonal directions for $\phi_{\mu}$ and $\phi_{\tau}$: 
\begin{itemize}
\vskip -.3cm
\item[-] the quadratic terms select the direction 
\beq
 \vev{\phi_{\mu}}= v_\mu (0,\cos\alpha,\sin\alpha),\vev{\phi_{\tau}}=v_\tau (0,\sin\alpha,-\cos\alpha)\,,\eeq 
\vskip -.3cm
\item[-] the trilinear term  is proportional to $ \cos 2\alpha$ and to maximize it we need $\alpha=\pi/2$ that gives the correct alignment to $\phi_\mu$ and $\phi_\tau$.
\vskip -.3cm
 \end{itemize}

Building $V^2_{soft}$ is a bit more \emph{ad hoc}:  we need to impose that $V^2_{soft} $ is invariant under one of  the following  transformations
\bea
i)&\quad&\phi_{1_1}\to  2 \phi_{1_3}  \quad \phi_{1_3}\to 1/2  \phi_{1_1}\,,\nn\\
ii)&\quad& \phi_{1_1}\to \phi_{1_2}\quad \phi_{1_2}\to \phi_{1_1}\,,\nn\\
iii)&\quad& \phi_1\to \phi_2
\eea
Clearly this transformations breaks explicitly $SU(3)$ and the $Z_2$ under which $\phi_2$ is odd. A possible $V^2_{soft}$ is  given by
\beq
m_1^2 |\phi_{1_1}-2 \phi_{1_3}|^2 +m_\pm^2  |\phi_{1_1}\pm \phi_{1_2}|^2+ m_{12}^2 ( \phi_1^\dag \phi_2+ H.c.)\,,
\eeq
By choosing correctly the sign of $m^2_1,m^2_-,m^2_{12}>0,m^2_+<0$ not to have tachyons  we get $\vev{\phi_1}$ and $\vev{\phi_2}$ orthogonal and along the right directions.

Finally $V^3_{soft}$ have to make massive the last 8 GBs: in order  not  to destroy the alignments provided by  $V^1_{soft}$ and  $V^2_{soft}$ it has to be subdominant with respect to them. It may have a form like
\beq
m_{\tau 1}^2 (\phi_\tau^\dag \phi_1)\,.
\eeq
In general   $V^3_{soft}$ leaves the freedom to preserve only one vev direction and   would slightly  disalign the  others. If only $\phi_1$--or $\phi_2$--is involved in $V^3_{soft}$  together with $\phi_{e,\mu,\tau}$  it is possible to disalign only the  triplets entering in the  charged lepton Yukawa lagrangian.  For example they would disalign according to 
\beq
\vev{\phi_{e,\mu,\tau}} + \epsilon_{e,\mu,\tau} (2,2,1)\,,
\eeq
giving rise to the corrections needed to generate a non trivial charge lepton mixing.

\section{Neutrino phenomenological analysis}
\label{anal}

\begin{figure}[h!]
\begin{center}
\includegraphics[width=4in]{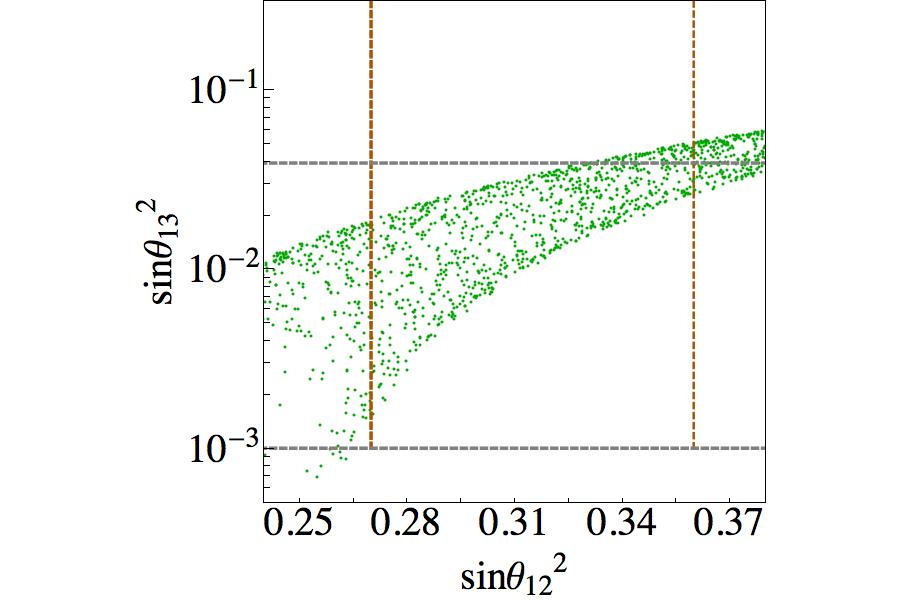}\\ \includegraphics[width=4in]{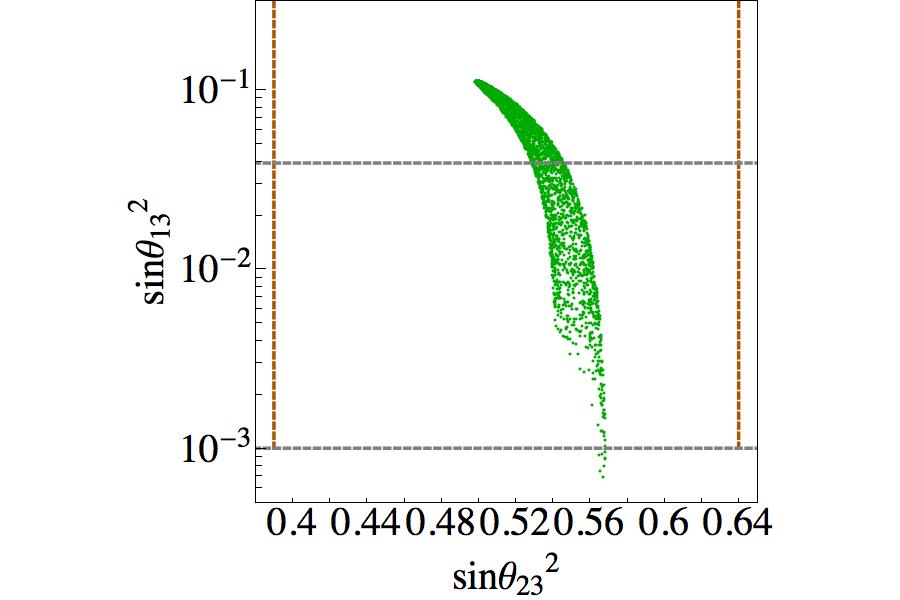} 
 \includegraphics[width=4in]{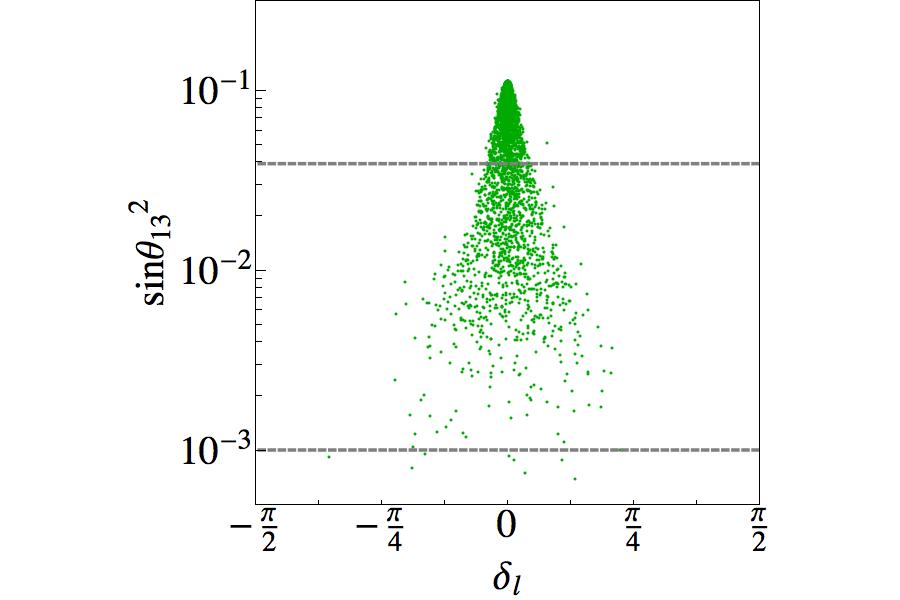}
\caption{\it The predictions for the lepton mixing parameters. From the upper to the lower panel we  plot the reactor  angle  versus the solar angle,  the atmospheric angle and the CP dirac phase $\delta_l$.  Vertical and horizontal lines bound the $3\sigma$ range for the corresponding angles.} \label{TPnum}
\end{center}
\end{figure}

Eq.(\ref{massnu}) and \eq{Ulep} give us all the informations to outline the neutrino phenomenology of the model discussed. For what concerns the spectrum only the  IH  case is  allowed  with vanishing $m_3$. In first approximation the predictions for the lepton angles coincide with those given in \eq{TPang} but a more accurate scan of the parameters space is performed by taking into account the complete  charged lepton mixing matrix obtained fitting the charged lepton masses. The result is showed in fig.\ref{TPnum}. From the upper to the lower panel we  plot the reactor  angle  versus the solar angle,  the atmospheric angle and the CP dirac phase $\delta_l$. The numerical scan confirms the parametric plot showed in the introduction. We have reported the $3\sigma$ range for the 3 angles according to the most recent analysis. There is a nice correlation between   the solar  and reactor angle: if  in the near future  there would be an improvement in the $3\sigma$ range precision of one of the two angles   we would automatically get an upper or/and lower predictions for the other.  On the other hand the model could be ruled out by an improvement on the precision for the atmospheric angle since it predicts $\sin\theta_{23}^2$ far from its central value (0.42). For what concerns the  CP Dirac phase our points clustered  in the range $0\pm\pi/4$. From \eq{expdelta} we see that $\delta_l\sim0$ is  the expected value from the analytical parametrization. Indeed  \eq{TPang} tell us that $\cos \delta \sim-1$  to fit  the solar angle.

\begin{figure}[h!]
\begin{center}
\includegraphics[width=4in]{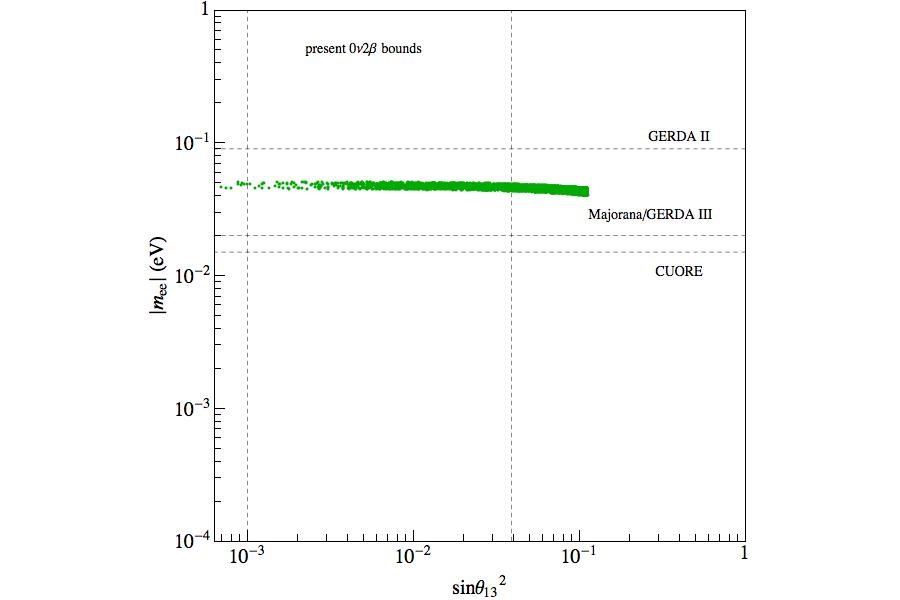}
\caption{\it The predictions for $m_{ee}$. Our IH spectrum is characterized by a vanishing $m_3$. As consequence we predict almost an exact value for  $m_{ee}\sim 45$ meV that is in the precision range of the forthcoming neutrino experiments. Vertical lines bound the $3\sigma$ range for $\sin \theta_{13}^2$. }
 \label{TPmee}
\end{center}
\end{figure}

To study neutrinoless double beta decay we consider the effective $0\nu\beta\beta$ parameter $m_{ee}$  defined as
\beq
m_{ee} = [U_{lep}\, \diag(m_1, \, m_2, \, 0) \, U_{lep} ]_{11}.
\eeq
In fig.\ref{TPmee} we plot $m_{ee}$ versus the  $\sin\theta_{13}^2$ since in our model the lightest neutrino mass is always vanishing. As consequence the model predicts almost an exact value for $m_{ee}\sim 45$ meV. The future experiments are expected to reach good sensitivities: $90$ meV \cite{gerda} (GERDA), $20$ meV \cite{majorana} (Majorana), $50$ meV \cite{supernemo} (SuperNEMO), $15$ meV \cite{cuore} (CUORE) and $24$ meV \cite{exo} (EXO). As a result the model may be tested in the next future.

\section{Conclusions}
\label{conc}

In this paper we have proposed a new mixing matrix for the neutrino mass matrix that we named the Tri-Permuting  (TP) mixing matrix. This pattern requires large corrections  both to the solar and to the reactor angles that thus are correlated in a new way orthogonal to other patterns proposed in literature such as the BM mixing one.  Not to affect neutrino masses and the atmospheric angle these corrections have to arise by  the charged lepton mixing matrix. We have build a full renormalizable model in which this scenario is realized. In the model proposed both neutrino and charged lepton get mass through a generalized see saw. The model is  characterized by a neutrino  IH spectrum with vanishing $m_3$. As consequence it is highly testable in the next future because it predicts an exact value for  the effective $0\nu\beta\beta$ parameter $m_{ee}\sim 45$ meV and could be ruled out by an improvement of precision for the atmospheric angle. At the same time it gives a nice correlation between solar and reactor angles that could be tested by future analysis. 

We have also  roughly discussed  the potential sketching the strategy to obtain the correct vacuum alignments and the correction to the charged lepton mass matrix needed to correct the TP mixing matrix.  A part from the neutrino sector the model  phenomenology  is deeply rich due to the presence of many new scalars and heavy fermions. A complete analysis of its phenomenology is postponed to a future work\cite{me}.

\section*{Aknowledgments}
I am grateful to S. Morisi for useful discussions and suggestions in the initial stage of this project.

\end{document}